\documentclass[useAMS,usenatbib]{mn2e}

\usepackage{epsfig}
\usepackage{longtable}
\usepackage{times}
\def \inte {{\em INTEGRAL}}

\def \sw {{\em Swift}}
\def \src {\mbox{IGR~J08408$-$4503}}
\def \igr {\mbox{IGR~J17544$-$2619}}
\def \xte {\mbox{XTE~J1739$-$302}}
\def \med {\mbox{IGR~J16479$-$4514}}

\def \hcm {\hbox {\ifmmode $ atom cm$^{-2}\else atom cm$^{-2}$\fi}}

\def \ATel {Astron.\ Tel.}
\def \apj {ApJ}
\def \apjl {ApJL}
\def \apjs {ApJS}
\def \aap {A\&A}
\def \aaps {A\&AS}

\def \pasj {PASJ}

\def \mnras {MNRAS}

\title[{\it Swift} observes outbursts from three SFXTs]{Supergiant Fast X--ray Transients in outburst: 
new {\it Swift} observations of \xte, \igr, and  \src}
\author[L. Sidoli et al.]{L.\ Sidoli$^{1}$, P.\ Romano$^{2}$,  L.\ Ducci$^{3,1}$, 
 A.\ Paizis$^{1}$, G.\ Cusumano$^{2}$, V.\ Mangano$^{2}$, 
\newauthor 
 H.A.\ Krimm$^{4,5}$, S.\ Vercellone$^{2}$, D.N.~Burrows$^{6}$, J.A.~Kennea$^{6}$, N.~Gehrels$^{5}$ \\
$^{1}$INAF, Istituto di Astrofisica Spaziale e Fisica Cosmica,
	Via E.\ Bassini 15,   I-20133 Milano,  Italy\\
$^{2}$INAF, Istituto di Astrofisica Spaziale e Fisica Cosmica,
        Via U.\ La Malfa 153, I-90146 Palermo, Italy\\
$^{3}$Dipartimento di Fisica e Matematica, Universit\`a dell'Insubria, Via Valleggio 11, I-22100 Como, Italy \\
$^{4}$Universities Space Research Association, Columbia, MD, USA \\
$^{5}$NASA/Goddard Space Flight Center, Greenbelt, MD 20771, USA\\
$^{6}$Department of Astronomy and Astrophysics, Pennsylvania State  University, University Park, PA 16802, USA\\
}

\begin{document}

\date{}

\pagerange{\pageref{firstpage}--\pageref{lastpage}} \pubyear{2009}

\maketitle

\label{firstpage}

\begin{abstract}
We report on new X--ray outbursts observed with {\it Swift} 
from three Supergiant Fast X--ray Transients (SFXTs): \xte, \igr, and \src.
\xte\ underwent a new outburst on  2008, August 13,
\igr\ on 2008, September 4, while \src\ on 2008, September 21.
While the \xte\ and \src\ bright emission triggered the \sw/Burst Alert Telescope, 
\igr\ did not, thus we could perform a spectral investigation
 only of the  spectrum below 10~keV.
The broad band spectra from \xte\ and \src\ were compatible with the X--ray spectral shape displayed 
during the previous flares.
A variable absorbing column density during the flare was observed in \xte\ for the first time. 
The broad-band spectrum of \src\
requires the presence of two distinct photon populations, a cold one 
($\sim$0.3\,keV) most likely from a thermal halo around the neutron star and a 
hotter one (1.4--1.8\,keV) from the accreting column.
The  outburst from \xte\ could be monitored with a very good sampling, thus revealing 
a shape which can be explained with a
second wind component in this SFXT, in analogy to what we have suggested in the periodic SFXT IGR~J11215--5952.
The outburst recurrence timescale in \igr\ during our monitoring campaign with Swift suggests a long orbital
period of $\sim$150~days (in an highly eccentric orbit), compatible with what previously observed with INTEGRAL.
\end{abstract}


	\section{Introduction\label{sfxt4:intro}}

The discovery of a new class of Galactic bright X--ray transients, 
composed of a compact object and an OB supergiant companion, the 
so-called Supergiant Fast X--ray Transients (SFXTs), is one of the
most intriguing results obtained by the \inte\ satellite
(\citealt{Sguera2005}, \citealt{Negueruela2006}).
Since its  launch in October 2002, the Galactic plane monitoring performed with
\inte/IBIS led to the discovery of several new sources  \citep{Bird2007}, some of which 
were characterized by short flares reaching 10$^{36}$--10$^{37}$~erg~s$^{-1}$,
and later optically associated
with blue supergiant stars (e.g. \citealt{Masetti2006}).
A couple of members of the class were discovered years before 2002, with
other satellites, and later re-discovered with \inte\ and  firmly 
classified as SFXTs:
XTE~J1739--302 \citep{Smith1998:17391-3021}, now considered the prototype of the class, 
and the X--ray pulsar AX~J1841.0--0536 \citep{Bamba2001}.
SFXTs display a broad band spectral shape similar 
to that of accreting X--ray pulsars \citep{Walter2006}, and
a high dynamic range in X--rays (up to four or five orders of magnitude)
down to a quiescent luminosity at $\sim$10$^{32}$~erg~s$^{-1}$ \citep{zand2005}.
Although X--ray pulsations have not been discovered, to date, in all
the members of the class, 
the spectral similarity with the accreting X--ray pulsars suggests that all, or
at least most of them, host neutron stars. 
Their peculiar transient behaviour is still waiting for a convincing theoretical explanation, 
although all the physical mechanisms proposed to date are mainly related 
to the structure of the supergiant wind  and/or to the properties of the accreting neutron star
(see \citealt{Sidoli2008:cospar} and references therein for a recent review).

The first systematic  monitoring of the X--ray activity of SFXTs
has been performed with \sw, during a campaign (still in progress) 
which started in October 2007, with the main aim
of studying the long-term properties of a sample of SFXTs: XTE~J1739--302/IGR~J17391--3021, 
IGR~J17544--2619, IGR~J16479--4514
and AX~J1841.0--0536/IGR~J18410--0535 (\citealt{Sidoli2008:sfxts_paperI}, hereafter Paper~I). 

One of the important results of these \sw\ observations is the
discovery that SFXTs do not spend most of the time in quiescence, as previously thought, but in an
intermediate level of emission at around  10$^{33}$--10$^{34}$~erg~s$^{-1}$,
displaying a hard X--ray spectrum and a frequent low intensity flaring activity,
with a dynamic range of more than one order of magnitude (Paper~I). 

During the \sw\ monitoring, about 1--2 bright outbursts per year per source have been caught
(\med, \citealt{Romano2008:sfxts_paperII}, hereafter Paper~II; 
\igr\ and \xte, 
\citealt{Sidoli2009:sfxts_paperIII}, hereafter Paper~III), 
leading to the study of the first broad band
spectrum from these sources, observed simultaneously from 0.3 to 60 keV.
The results of a \sw\ monitoring of a multi-flaring X--ray activity in July 2008 
from a fifth SFXT, \src, has been 
reported by \citet{Romano2009:sfxts_paper08408}.

In this paper we report on the latest three outbursts caught by \sw\  
from three SFXTs, already announced to the scientific community:  
XTE~J1739--302 (2008, August 13; \citealt{Romano2008:atel1659}), 
IGR~J17544--2619 (2008, September 4; \citealt{Romano2008:Atel1697}). 
\src\ (2008, September 21; \citealt{Mangano2008:atel1727}).

\begin{table*}
 \begin{center}
 \caption{Observation log.\label{sfxt4:tab:observations} }
 \begin{tabular}{lllllr}
 \hline
 \hline
 \noalign{\smallskip}
Source &  Sequence & Instrument & Start time (UT) & End time (UT) &  Exposure \\ 
  & & /Mode & (yyyy-mm-dd hh:mm:ss) & (yyyy-mm-dd hh:mm:ss) & (s) \\
  \noalign{\smallskip}
 \hline
 \noalign{\smallskip}
XTE~J1739$-$302    & 00319963000     &       BAT/event &       2008-08-13 23:47:14     &       2008-08-14 00:05:22     & 1088  \\ 
                   & 00319963000     &       XRT/WT    &       2008-08-13 23:55:55     &       2008-08-14 00:29:09     &       1688  \\  
                   & 00319963000     &       XRT/PC    &       2008-08-14 00:03:19     &       2008-08-14 00:04:34     &       75      \\
                   & 00319964000     &       BAT/event &       2008-08-14 00:08:56     &       2008-08-14 00:28:58     & 728   \\
                   &00030987070	&XRT/WT   &	  2008-08-14 01:23:10	  &	  2008-08-14 13:03:50	  &	  1207    \\
                   &00030987070	&XRT/PC   &	  2008-08-14 04:36:11	  &	  2008-08-14 13:17:16	  &	  10714   \\
                   &00030987071	&XRT/PC   &	  2008-08-15 00:00:34	  &	  2008-08-15 00:16:57	  &	  983	  \\
                   &00030987072	&XRT/PC   &	  2008-08-16 03:21:09	  &	  2008-08-16 05:08:57	  &	  1379    \\
                   &00030987073	&BAT/PC   &	  2008-08-17 01:44:09	  &	  2008-08-17 02:01:57	  &	  1068    \\
                   &00030987074	&XRT/PC   &	  2008-08-18 13:04:08	  &	  2008-08-18 13:21:56	  &	  1068    \\
                   &00030987075	&XRT/PC   &	  2008-08-19 13:09:42	  &	  2008-08-19 13:27:58	  &	  1095    \\
                   &00030987076	&XRT/PC   &	  2008-08-20 05:14:06	  &	  2008-08-20 05:31:56	  &	  1071    \\
                   &00030987077	&XRT/PC   &	  2008-08-21 00:30:30	  &	  2008-08-21 15:08:55	  &	  1173    \\
                   &00030987078	&XRT/PC   &	  2008-08-22 00:36:19	  &	  2008-08-22 02:21:55	  &	  1093    \\
                   &00030987079	&XRT/PC   &	  2008-08-23 03:55:11	  &	  2008-08-23 05:41:56	  &	  1229    \\
                   &00030987080	&XRT/PC   &	  2008-08-24 21:44:19	  &	  2008-08-24 23:31:57	  &	  1332    \\
                   &00030987081	&XRT/PC   &	  2008-08-25 21:50:21	  &	  2008-08-25 23:35:56	  &	  1274    \\
IGR~J08408$-$4503  & 00325461000     &       BAT/event &       2008-09-21 07:54:11     &       2008-09-21 09:16:38     & 903   \\
                   & 00325461000     &       XRT/WT    &       2008-09-21 08:03:54     &       2008-09-21 14:05:06     &       1159    \\
                   & 00325461000     &       XRT/PC    &       2008-09-21 09:16:31     &       2008-09-21 14:24:59     &       4116    \\
                   & 00030707013     &       XRT/PC    &       2008-09-23 15:53:39     &       2008-09-23 17:36:57     &       1109    \\
                   & 00030707014     &       XRT/PC    &       2008-09-25 18:15:07     &       2008-09-25 19:57:54     &       1094    \\
                   & 00030707015     &       XRT/PC    &       2008-09-26 18:10:01     &       2008-09-26 18:26:37     &       995     \\
                   & 00030707016     &       XRT/PC    &       2008-09-27 16:39:46     &       2008-09-27 18:19:57     &       216     \\
                   & 00030707017     &       XRT/PC    &       2008-09-28 16:56:30     &       2008-09-28 17:01:26     &       295     \\
                   & 00030707018     &       XRT/PC    &       2008-09-29 10:36:24     &       2008-09-29 17:02:57     &       521     \\
IGR~J17544$-$2619  & 00035056061     &       XRT/WT    &       2008-09-04 00:12:40     &       2008-09-04 00:25:03     &       217     \\
                   & 00035056061     &       XRT/PC    &       2008-09-04 00:12:45     &       2008-09-04 00:26:55     &       632     \\
  \noalign{\smallskip}
  \hline
  \end{tabular}
  \end{center}
\end{table*}

\subsection{Previous observations of the three SFXTs}


\xte\ was discovered by $RXTE$ in August 1997 \citep{Smith1998:17391-3021}, 
reaching a peak flux of 3.6$\times$10$^{-9}$~erg cm$^{-2}$~s$^{-1}$   (2--25 keV).
The optical counterpart is an  O8~Iab(f) star \citep{Negueruela2006}
located at 2.7 kpc \citep{Rahoui2008}.
ASCA observations 
\citep{Sakano2002} allowed a constraint on the quiescent emission at a level of $<$1.1$\times$10$^{-12}$~erg cm$^{-2}$~s$^{-1}$.
The source showed  bright outbursts, reaching 300~mCrab, observed with IBIS/ISGRI in 2003 March, 
and 2004 March \citep{Sguera2006}. 
Other flares  observed with \inte\ have been reported by \citet{Walter2007} and \citet{Blay2008}. 
More recently, it triggered the {\it Swift} Burst Alert Telescope (BAT) on 2008 April 8,
when a  bright flare was caught, reaching an 
X--ray luminosity  of $\sim$3$\times$10$^{36}$~erg~s$^{-1}$ (0.5--100~keV; Paper~III).
Since October 2007 the source has been monitored with \sw/XRT $\sim$2--3 times a week, showing
a high variability in its flux even outside outbursts (Paper~I).


\igr\ was discovered  with \inte\ 
on 2003 September 17  during a short flare reaching 
160~mCrab (18--25~keV; \citealt{Sunyaev2003}). 
A $Chandra$ observation performed in 2004 caught both the quiescence level and the onset of
an outburst \citep{zand2005}, translating into a dynamic
range as large as 10$^{4}$.
The optical counterpart is an  O9Ib star \citep{Pellizza2006}
located at 3.6 kpc \citep{Rahoui2008}.
Several more bright flares have been observed with \inte\
in 2003, 2004 and 2005 (\citealt{Grebenev2003:17544-2619}, \citealt{Grebenev2004:17544-2619},
\citealt{Sguera2006},  \citealt{Walter2007},  \citealt{Kuulkers2007}, and \citealt{Ducci2008}),
Two  outbursts were detected with the {\it Swift} satellite,
on 2007 November 8  \citep{Krimm2007:ATel1265}
and on 2008 March 31 \citep{Sidoli2008:atel1454}, 144 days apart. 
Fainter activity at a level of 30--40 mCrab (20--60 keV) from \igr\ was 
also observed on  2007 September 21, with IBIS/ISGRI on-board \inte\
\citep{Kuulkers2007}.
\igr\ is one of the four SFXTs we have been monitoring with \sw/XRT since October 2007 (Paper~I).


\src\ was discovered on 2006 May 15, when a  bright 
flare  reaching a peak flux of 250 mCrab in the 
20--40 keV energy band was caught by \inte\ \citep{Gotz2006:08408-4503discovery}. 
Analysis of archival \inte\ observations of the source field showed that 
\src\ was previously active on 2003 July 1 \citep{Mereghetti:08408-4503}.
A refined position with \sw/XRT \citep{Kennea2006:08408-4503}
allowed to associate the source with a O8.5Ib(f) supergiant star, HD~74194, 
at a distance of about 3\,kpc \citep{Masetti2006:08408-4503}.
Three additional flares observed with \inte\ and \sw\ were studied by \citet{Gotz2007:08408-4503}.
A new outburst from \src\ was caught on 2008 July 5 by \sw/BAT 
and then followed up at softer energies with \sw/XRT \citep{Romano2009:sfxts_paper08408}.
In that occasion the source displayed a multiple flaring activity (the XRT lightcurve 
showed three bright flares in excess of 10~s$^{-1}$). 
The properties of the flares and of 
the times of the outbursts suggested an orbital period of $\sim$35~days \citep{Romano2009:sfxts_paper08408}.

 	 \section{Observations and Data Reduction\label{sfxt4:dataredu}}

As a response to a first BAT trigger from \xte\ on 2008-08-13 at 23:49:17 UT 
(image trigger 319963), \sw\ executed an immediate slew and was on target 
in $\sim$390\,s; a second trigger (319964) occurred while \xte\ was in the XRT 
field of view on 2008-08-14 at 00:12:53 UT. 
%
The bright flare of \igr\ was instead discovered as part of the yearly monitoring 
with \sw/XRT, starting on 2008-09-04 at about 00:19:00 UT. 
The \sw/BAT did not trigger on it.  
%
\src\ triggered the BAT on 2008-09-21 at 07:55:08 UT (image trigger 325461). \sw\ slewed immediately and 
the NFI were on target in $\sim$147\,s.

Table~\ref{sfxt4:tab:observations} reports the log of the \sw\ observations of the 
outbursts used for this work. 
The XRT data were processed with standard procedures 
({\sc xrtpipeline} v0.12.1), filtering and screening criteria by using 
{\sc FTOOLS} in the {\sc Heasoft} package (v.6.6.1). 
We considered both WT and PC data, 
and selected event grades 0--2 and 0--12, respectively 
(\citealt{Burrows2005:XRTmnras}).
When appropriate we corrected for pile-up by determining the size of the 
PSF core affected by comparing the observed and nominal PSF \citep{vaughan2006:050315mnras},
and excluding from the analysis all the events that fell within that
region. Background events were extracted in source-free annular regions, centred on the
source. 
Ancillary response files were generated with {\sc xrtmkarf},
and they account for different extraction regions, vignetting, and
PSF corrections. We used the latest spectral redistribution matrices
(v011) in CALDB.

The  BAT data were collected in event mode for several hundred seconds after 
the triggers of \xte\ and \src, as detailed below, while 
\igr\ was not detected. 
The BAT data were analysed using the standard BAT analysis 
software distributed within {\sc FTOOLS}.
BAT mask-weighted spectra were extracted over the time
intervals simultaneous with XRT data when possible and 
response matrices were generated with {\sc batdrmgen}.
For our spectral fitting (XSPEC v11.3.2)
we applied an energy-dependent
systematic error. 

All quoted uncertainties are given at 90\% confidence level for 
one interesting parameter unless otherwise stated. 
The spectral indices are parameterized as  
$F_{\nu} \propto \nu^{-\alpha}$, 
where $F_{\nu}$ (erg cm$^{-2}$ s$^{-1}$ Hz$^{-1}$) is the 
flux density as a function of frequency $\nu$; 
we adopt $\Gamma = \alpha +1$ as the photon index, 
$N(E) \propto E^{-\Gamma}$ (ph cm$^{-2}$ s$^{-1}$ keV$^{-1}$). 
Times in the light curves and the text are referred to their respective 
BAT triggers with the exception of \igr\ 
which did not trigger the BAT, thus the start time was set
at the beginning of the observation.

	\section{Analysis and Results\label{sfxt4:res}}

\subsection{\xte\label{broadxte}}

\begin{figure}
\begin{center}
\includegraphics*[angle=270,scale=0.35]{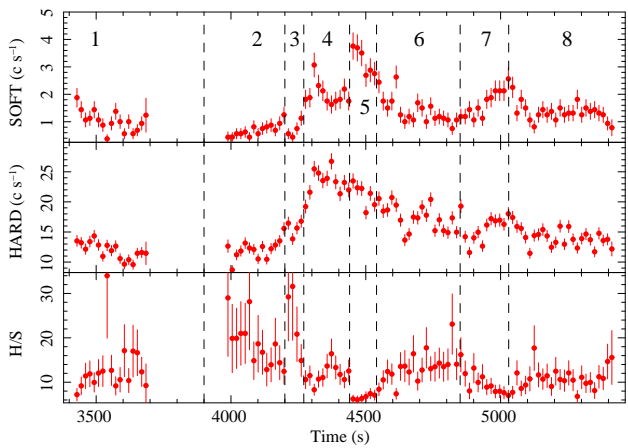}
\end{center}
\caption{\sw/XRT (WT data) light curves of \xte\ in two energy ranges, 0.3-2 keV (SOFT, upper panel) 
and 2--10~keV (HARD, middle panel), together with their hardness ratio (lower panel).
Vertical dashed lines mark the eight spectra selected for the time resolved spectroscopy.
}
\label{lsfig:xte:hr}
\end{figure}

\begin{figure}
\begin{center}
\includegraphics*[angle=270,scale=0.35]{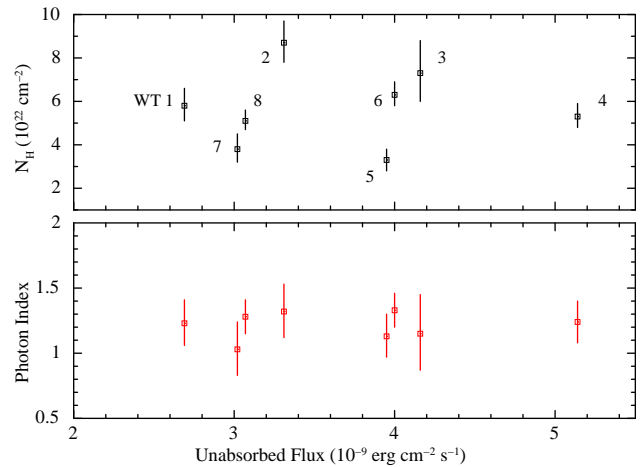}
\end{center}
\caption{\sw/XRT (WT data) time resolved spectroscopy  
of \xte: spectral results of the absorbed power law fit (reported in Table~\ref{tab:xte_8bins}).
Numers mark the eight WT spectra, as in Table~\ref{tab:xte_8bins}.
}
\label{lsfig:xte:nhgamma}
\end{figure}

The complete light curve of the bright flaring of \xte\ as observed with \sw/XRT 
on 2008 August 13 is reported in Fig.~\ref{lsfig:duration}c, while
the first part ($\sim$2000~s) of the observation (WT data, observation 00319963000), is expanded in 
Fig.~\ref{lsfig:xte:hr},
where a soft (below 2 keV) and a hard (above 2 keV) light curves are reported, together
with their hardness ratio. 
Since the source hardness appears to be variable, we 
performed a time resolved spectroscopy extracting eight XRT/WT spectra as shown
in  Fig.~\ref{lsfig:xte:hr}.
These spectra could be adequately fitted both with an absorbed power law model 
and with an absorbed single blackbody (see Table~\ref{tab:xte_8bins} for the results, spectra from WT~1 to WT~8).
There is a clear time variability of the absorbing column density (by a factor of $\sim$3), 
whereas the spectral
shape (photon index, $\Gamma$, or the blackbody temperature, kT$_{\rm bb}$) remain constant, 
within the uncertainties (see Fig.~\ref{lsfig:xte:nhgamma}).
In particular, spectra WT~3 and WT~5 are the hardest and
the softest one, respectively, thus demonstrating that the hardness ratio variability 
in Fig.~\ref{lsfig:xte:hr} is due to a variable absorption into the line of sight.
As  final tests, we fixed the photon index $\Gamma$=1.15 (average value) and then re-fitted the eight spectra.
This still resulted in a variable absorbing column density.
We then fixed the absorbing column density to an mean value of 5$\times$10$^{22}$ cm$^{-2}$
and refitted the spectra. Those spectra where the absorption were previously found to be
very different from this mean value, resulted in unacceptable fits.

\begin{figure}
\begin{center}
\includegraphics*[angle=270,scale=0.35]{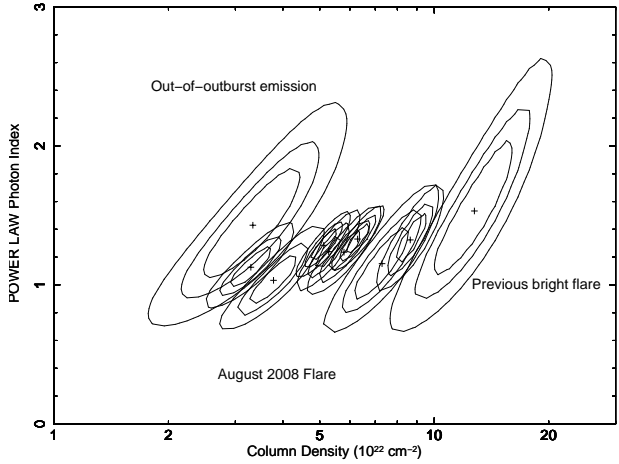}
\end{center}
\caption{\sw/XRT (WT data) time resolved spectroscopy  (eight spectra reported in Table~\ref{tab:xte_8bins}) 
of \xte\ during the new outburst observed 
on 2008, August 13 (absorbed power law) compared with the spectral paramenters 
derived during the previous bright flare observed with \sw\ (discussed in Paper~III; large contours on the right).
The power law parameters of the total spectrum of the out-of-outburst emission  (Paper~I) are also shown (large contours on the left).
68\%, 90\% and 99\% confidence level contours are shown.
}
\label{lsfig:xte:cont}
\end{figure}

Figure~\ref{lsfig:xte:cont} shows the comparison between the time resolved spectroscopy 
of the XRT/WT data (Table~\ref{tab:xte_8bins}) of the August 2008 outburst 
and two more spectral analyses: the out-of-outburst emission 
(Paper~I) and the spectroscopy of a previous  flare from \xte\ (Paper~III).
There is no evidence for a spectral change with the source flux, nor for a correlation
of the absorbing column density with the source flux.
The absorption is higher during the rising phase of the bright flare.

 \begin{table*}
 \begin{center}
 \caption{Time resolved spectroscopy of \xte\ (XRT data).}
 \label{tab:xte_8bins}
 \begin{tabular}{llllrrrrr}
 \hline
 \hline
 \noalign{\smallskip}
 Spectrum  & Time &  Model & $N_{\rm H}$ 	   &  $\Gamma$                 &   kT$_{bb}$  &  Flux$^{\mathrm{a}}$  & R$_{bb}$          & $\chi^{2}_{\nu}$ /d.o.f. \\
           & (s since trigger)     &        & ($10^{22}$ cm$^{-2}$) &                           &    (keV)     &                     &   (km)$^{\mathrm{b}}$ &         \\
  \noalign{\smallskip}
 \hline
 \noalign{\smallskip}
  WT 1  & 397--656   & Pow &  $5.8_{-0.7}^{+0.8}$ 	   &  $1.23_{-0.17}^{+0.18}$ &                       & 2.69     &                     & $1.191/139$\\
        &            & BB &  $3.1_{-0.4}^{+0.5}$ 	   &                       & $1.88_{-0.11}^{+0.12}$  &          &  $1.1_{-0.1}^{+0.1}$  & $1.044/139$\\
  WT 2  & 961--1170  &  Pow  &  $8.7_{-0.9}^{+1.0}$    &  $1.32_{-0.20}^{+0.21}$ &                       & 3.31     &                     & $1.038/112$\\
        &            & BB &  $5.4_{-0.6}^{+0.7}$       &                       & $1.93_{-0.14}^{+0.16}$  &          &  $1.2_{-0.1}^{+0.2}$  & $1.080/112$\\
  WT 3  & 1170--1255 &  Pow&  $7.3_{-1.3}^{+1.5}$      &  $1.15_{-0.28}^{+0.30}$ &                       & 4.16     &                     & $0.873/66$\\
        &            & BB &  $4.4_{-0.8}^{+0.9}$       &                       & $1.98_{-0.19}^{+0.23}$  &          &  $1.3_{-0.2}^{+0.3}$  & $0.809/66$\\
  WT 4  & 1255--1407 &  Pow &  $5.3_{-0.5}^{+0.6}$     &  $1.24_{-0.16}^{+0.16}$ &                       & 5.14     &                     & $0.961/159$\\
        &            & BB &  $2.8_{-0.3}^{+0.4}$       &                       & $1.82_{-0.10}^{+0.11}$  &          &  $1.7_{-0.1}^{+0.2}$  & $0.910/159$\\
  WT 5  & 1410--1520 &  Pow &  $3.3_{-0.5}^{+0.5}$     &  $1.13_{-0.16}^{+0.17}$ &                       & 3.95     &                     &  $1.074/114$ \\
        &            & BB &  $1.4_{-0.2}^{+0.3}$       &                       & $1.80_{-0.11}^{+0.12}$  &          &  $1.5_{-0.1}^{+0.2}$  &  $0.846/114$ \\
  WT 6  & 1522--1876 & Pow &  $6.3_{-0.5}^{+0.6}$      &  $1.33_{-0.13}^{+0.13}$ &                       & 4.00     &                     &  $1.241/243$ \\
        &            & BB &  $3.4_{-0.3}^{+0.3}$       &                       & $1.83_{-0.08}^{+0.08}$  &          &  $1.4_{-0.1}^{+0.1}$  &  $1.047/243$ \\
  WT 7  & 1877--2005 & Pow &  $3.8_{-0.6}^{+0.7}$      &  $1.03_{-0.20}^{+0.21}$ &                       & 3.02     &                     &  $1.345/98$\\
        &            & BB &  $1.8_{-0.3}^{+0.4}$       &                       & $1.90_{-0.14}^{+0.16}$  &          &  $1.2_{-0.1}^{+0.2}$  &  $1.227/98$\\
  WT 8  & 2006--2390 & Pow &  $5.1_{-0.4}^{+0.5}$      &  $1.28_{-0.13}^{+0.13}$ &                       & 3.07     &                     &  $1.083/229$ \\
        &            & BB &  $2.6_{-0.3}^{+0.3}$       &                       & $1.82_{-0.08}^{+0.09}$  &          &  $1.3_{-0.1}^{+0.1}$  &  $0.969/229$	\\
  WT 9$^{\mathrm{c}}$  & 5631--47671& Pow &  $3.6_{-0.3}^{+0.4}$     & $1.1_{-0.1}^{+0.1}$ &      &  1.02  &                     &  $1.396/259$ \\
        &            & BB               &  $1.5_{-0.2}^{+0.2}$     &           & $1.89_{-0.07}^{+0.08}$  &          & $0.72_{-0.04}^{+0.05}$   &   $1.128/259$  \\
  PC 1$^{\mathrm{c}}$  &17212--48475& Pow &  $3.0_{-0.4}^{+0.4}$ &   $1.26_{-0.16}^{+0.17}$  &            &  0.23  &                     &  $1.095/113$ \\
        &            & BB &  $1.3_{-0.2}^{+0.2}$      &                       &    $1.64_{-0.09}^{+0.10}$ &          &  $0.43_{-0.03}^{+0.04}$  & $1.082/113$ \\
  \noalign{\smallskip}
  \hline
  \end{tabular}
  \end{center}
  \begin{list}{}{}
  \item[$^{\mathrm{a}}$]{Unabsorbed 1--10\,keV flux in units of $10^{-9}$ erg cm$^{-2}$ s$^{-1}$.}
  \item[$^{\mathrm{b}}$]{Assuming a distance of 2.7~kpc.}
  \item[$^{\mathrm{c}}$]{Observation 00030987070.}
 \end{list}
  \end{table*}


A high energy spectrum (BAT) was also available, but only simultaneously to the XRT/WT spectrum n.~1.
A joint fit of XRT/WT and BAT spectra was performed 
including constant factors to allow for normalization uncertainties between the two
instruments (always constrained to be within their usual ranges).
A single power law is unable to describe the broad band spectrum.
We then tried models usually adopted to describe the X--ray emission from accreting pulsars,
resulting in the spectral parameters listed in Table~\ref{tab:xte:spec}: 
a cutoff power law (E$^{-\Gamma}$$e^{(-E/E_{\rm cut})}$), 
and two kind of Comptonization models. 
The best deconvolution of the broad band spectrum 
has been obtained with these latter models:     
a Comptonization of seed photons (with a temperature kT$_0$) in a hot plasma (with electron temperature kT$_{\rm e}$) 
as described by {\sc compTT} in {\sc xspec} \citep{Titarchuk1994},
or by {\sc bmc} \citep{TMK1996}.

The {\sc bmc} model is the sum of a blackbody (BB) plus its Comptonization, the 
latter obtained as a consistent convolution of the blackbody itself  with 
the Green's function of the Compton corona. The {\sc bmc} model is not limited 
to the thermal Comptonization case (as e.g. {\sc compTT}) and accounts also for 
dynamical (bulk) Comptonization due to the converging flow. 
Similarly to the ordinary {\sc bbody xspec} model, the normalization of {\sc bmc} is 
the ratio of the source luminosity to the square of the distance 
(in units of 10\,kpc).
The free parameters of the {\sc bmc} model (apart from the 
normalization) are the black-body (BB) colour temperature, $kT_{\rm BB}$, the spectral 
index $\alpha$  
and the logarithm of the illuminating factor A, $\log A$. 
The parameter $\alpha$ indicates the overall Comptonization efficiency 
related to an observable 
quantity in the photon spectrum of the data. The lower $\alpha$, the higher 
the efficiency
i.e. the higher the energy transfer from the hot electrons to the soft 
seed photons. 
The  $\log A$ parameter is an indication of the fraction of the 
up-scattered BB photons with 
respect to the BB seed photons directly visibile. In the extreme cases, 
the seed photons can be completely embedded 
in the Comptonizing cloud (none directly visible, $A\gg1$, e.g. $\log A=8$) 
or there is no coverage by 
the Compton cloud  (A $\ll$1, e.g. $\log A=-8$) and we observe directly the 
seed photon spectrum 
(equivalent to a simple BB, with no Comptonization). 

\begin{figure}
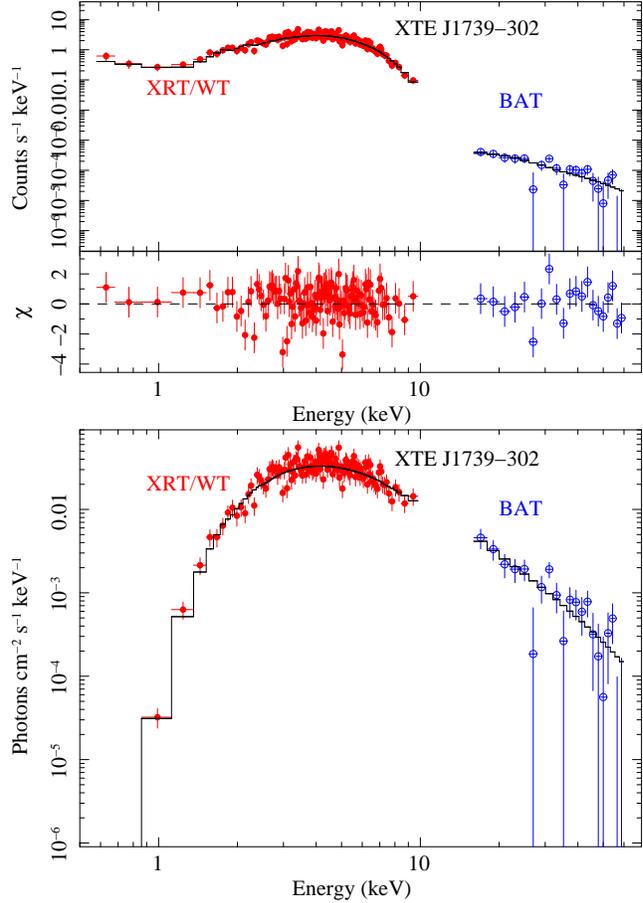

\begin{center}
\includegraphics*[angle=270,scale=0.35]{xte_ldadelchi_comptt_2.ps}
\vspace{1cm}
\includegraphics*[angle=270,scale=0.35]{xte_ufs_comptt_2.ps}
\end{center}
\caption{\xte\ broad band spectrum (XRT/WT together with BAT data) fitted with a Comptonization model
({\sc compTT} in {\sc xspec}).
}
\label{lsfig:xte:comptt}
\end{figure}

The \xte\ broad band spectrum fitted with the {\sc compTT} model is
shown in Fig.~\ref{lsfig:xte:comptt}. 
The estimated X--ray luminosity during the flare is 3.8$\times$10$^{36}$~erg~s$^{-1}$ (0.1--100~keV),
at a source distance of 2.7~kpc. 

As can be seen in Table~\ref{tab:xte:spec}, the parameters describing the properties of 
the Comptonizing corona (be they the temperature and optical depth in 
{\sc compTT} or the illumination parameter logA in {\sc bmc}) are not constrained. 
This is expected given the poor statistics of the high energy part of the 
spectrum. Nevertheless, the models applied do give a first order 
description of the physical processes involved in this system (see 
Discussion).

 \begin{table*}
 \begin{center}
 \caption{Spectral fits of simultaneous XRT and BAT data of \xte.}
 \label{tab:xte:spec}
 \begin{tabular}{lrrrrrr}
 \hline
 \hline
 \noalign{\smallskip}
Model  &  	   &                &   Parameters &          &                        &           \\
 \hline
\sc{cutoffpl}     &  $N_{\rm H}$$^{\mathrm{a}}$    &     $\Gamma$          &   E$_{\rm cut}$$^{\mathrm{b}}$  &     &  Flux$^{\mathrm{c}}$  & $\chi^{2}_{\nu}$/d.o.f. \\
                  &  $5.5_{-0.8}^{+0.8}$ 	&  $0.87_{-0.28}^{+0.27}$ &     $15_{-4}^{+6}$         &     &    6.6             & $1.166/161$\\
\sc{bmc}         &  $N_{\rm H}$$^{\mathrm{a}}$ &    kT$_{\rm BB}$$^{\mathrm{b, d}}$     &   $\alpha$$^{\mathrm{d}}$            & $\log(A)$$^{\mathrm{d}}$ &  Flux$^{\mathrm{c}}$& $\chi^{2}_{\nu}$/d.o.f. \\
                 &  $3.4_{-0.5}^{+0.6}$ &  $1.6_{-0.3}^{+0.2}$ & $1.3_{-0.4}^{+0.4}$  & $0.5_{-0.6}^{+7.5}$  &    4.9            & $1.080/160$\\
\sc{compTT}$^{\mathrm{e}}$ &  $N_{\rm H}$$^{\mathrm{a}}$ &  kT$_0$$^{\mathrm{b}}$  & kT$_{\rm e}$$^{\mathrm{b}}$ & $\tau$  &  Flux$^{\mathrm{c}}$& $\chi^{2}_{\nu}$/d.o.f. \\
            &  $2.8_{-0.5}^{+0.6}$       &  $1.35_{-0.17}^{+0.12}$ & $>9$                     & $4.3_{-4.0}^{+3.6}$     &    4.6            & $1.061/160$\\
 \noalign{\smallskip}
  \hline
  \end{tabular}
  \end{center}
  \begin{list}{}{} 
 \item[$^{\mathrm{a}}$]{Absorbing column density is in units of $10^{22}$ cm$^{-2}$.}
   \item[$^{\mathrm{b}}$]{High energy cutoff (E$_{\rm cut}$), electron temperature (kT$_{\rm e}$), 
seed photons temperature (kT$_0$) and  the blackbody color temperature kT$_{\rm BB}$ are all in units of keV.}
  \item[$^{\mathrm{c}}$]{Unabsorbed 0.1--100\,keV flux is in units of $10^{-9}$ erg cm$^{-2}$ s$^{-1}$.}
 \item[$^{\mathrm{d}}$]{kT$_{\rm BB}$ is the blackbody color temperature of the seed photons, $\alpha$ 
is the spectral index and Log(A) is the illumination parameter (see Sect.~\ref{broadxte} for details).}
 \item[$^{\mathrm{e}}$]{Assuming a spherical geometry.}
  \end{list}
  \end{table*}

\subsection{\igr}

\begin{figure}
\begin{center}
\includegraphics*[angle=270,scale=0.35]{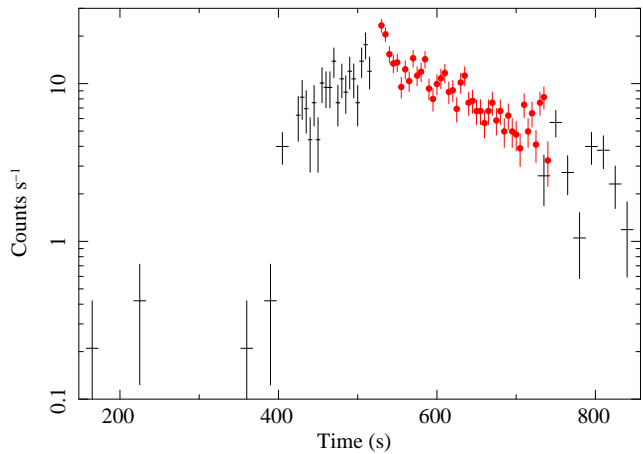}
\end{center}
\caption{\sw/XRT light curve of \igr\ during the outburst observed  
on 2008, September 4. Crosses mark the WT spectrum, circles refer to PC data. 
}
\label{lsfig:igr_lcv}
\end{figure}

A new bright flare from \igr\ was caught on 2008 September 4  with \sw, and it 
was preceded by intense activity in the previous few days  
observed during the \inte\ Galactic bulge monitoring program \citep{Romano2008:Atel1697},
reaching about 50 mCrab (18--40 keV) on 2008 August 30.
The flare was caught by \sw/XRT, only thanks to the on-going monitoring campaign
just targeted on the source, while
\sw/BAT did not trigger on it. 
The XRT light curve (Fig.~\ref{lsfig:igr_lcv}) shows a peak exceeding 20~s$^{-1}$, brighter 
than the previous one observed with \sw\ on March 31, at 20:53:27 UT (Paper~III).

\begin{figure}
\begin{center}
\includegraphics*[angle=270,scale=0.35]{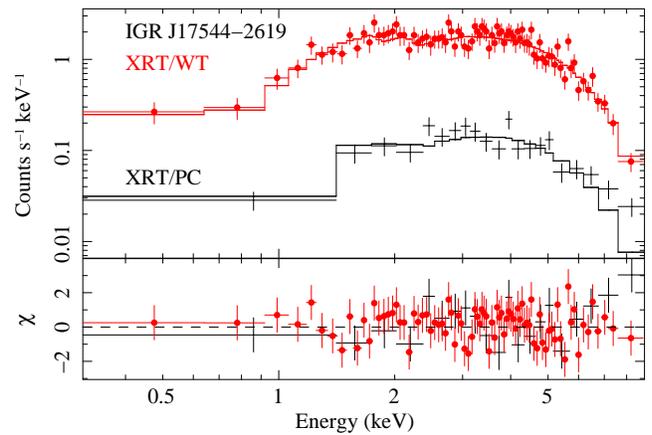}
\end{center}
\caption{\igr\ spectra (PC plus WT data) fitted with an absorbed blackbody 
(see Table~\ref{tab:igr:spec} for the parameters), together with the residuals
in units of standard deviations. Crosses mark the WT spectrum, circles refer to PC data (same as Fig.~\ref{lsfig:igr_lcv}). 
}
\label{lsfig:igr:spec}
\end{figure}

The total PC spectrum resulted in an exposure time of 632~s and a source net count rate
of 0.67$\pm{0.03}$~s$^{-1}$, while the WT data, extracted with a net exposure of 217~s, resulted
in an average count rate of 8.2$\pm{0.2}$~s$^{-1}$. 
Fitting the two spectra separately with simple models (an absorbed power law or a blackbody)
resulted in the parameters listed in Table~\ref{tab:igr:spec}. 
A black body is a better fit to the WT data than a single power law, which produces systematic positive residuals
around 1~keV. 
The resulting black body radius at an assumed source 
distance of 3.6 kpc is R$_{\rm bb}$=$1.35_{-0.12}^{+0.15}$~km.
More complex models are not required by the data.
We also fit together PC and WT spectra, adopting free 
normalization constant factors between the two spectra. 
The best fit obtained with a blackbody model of the joint PC plus WT 
data is reported in Fig.~\ref{lsfig:igr:spec}.

 \begin{table*}
 \begin{center}
 \caption{\igr\ spectral fits of XRT data.}
 \label{tab:igr:spec}
 \begin{tabular}{lllrrrrr}
 \hline
 \hline
 \noalign{\smallskip}
 Spectrum  &   Model & $N_{\rm H}$ 	   &  $\Gamma$                 &   kT$_{bb}$  &  Flux$^{\mathrm{a}}$  & R$_{bb}$             & $\chi^{2}_{\nu}$/d.o.f. \\
           &         & ($10^{22}$ cm$^{-2}$) &                           &    (keV)     &                     &   (km)$^{\mathrm{b}}$ &         \\
  \noalign{\smallskip}
 \hline
 \noalign{\smallskip}
  Total WT  &  Pow  &  $1.8_{-0.3}^{+0.4}$    &  $1.28_{-0.19}^{+0.18}$ &                       & 0.97     &                     & $1.170/82$\\
            &  BB   &  $0.52_{-0.14}^{+0.17}$ &                       & $1.51_{-0.09}^{+0.10}$  &          &  $1.35_{-0.12}^{+0.15}$  & $0.808/82$\\
 Total PC   &  Pow  &  $1.3_{-0.5}^{+0.8}$    &  $0.76_{-0.35}^{+0.40}$ &                       & 0.27     &                     & $1.290/18$\\
            &  BB   &  $0.42_{-0.26}^{+0.38}$    &                       & $1.87_{-0.29}^{+0.40}$  &          &  $0.51_{-0.10}^{+0.17}$  & $1.080/18$\\
\hline
joint fit WT + PC    & Pow  &  $1.8_{-0.9}^{+1.0}$    &  $1.20_{-0.16}^{+0.17}$ &                       &      &                     & $1.227/102$\\
                    & BB   &  $0.52_{-0.13}^{+0.15}$    &                       & $1.55_{-0.09}^{+0.09}$  &          &   & $0.979/102$\\
\noalign{\smallskip}
  \hline
  \end{tabular}
  \end{center}
  \begin{list}{}{}
  \item[$^{\mathrm{a}}$]{Unabsorbed 1--10\,keV flux in units of $10^{-9}$ erg cm$^{-2}$ s$^{-1}$.}
  \item[$^{\mathrm{b}}$]{Assuming a distance of 3.6~kpc.}
  \end{list}
  \end{table*}

\subsection{\src}
\label{bmc}

\begin{figure}
\begin{center}
\includegraphics*[angle=270,scale=0.35]{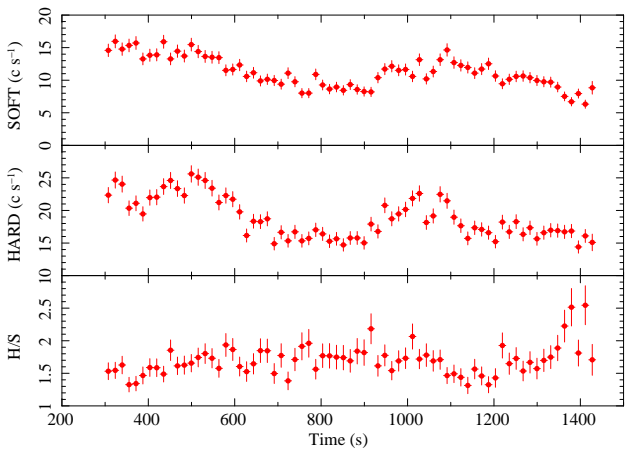}
\end{center}
\caption{\sw/XRT light curves (WT data) of \src\ during the 2008, September 21 outburst 
(time is in seconds since the BAT trigger). The bin time is 16 s. The soft and hard energy ranges
are below and above 2 keV, respectively. The lower panel shows their hardness ratio.
}
\label{lsfig:src_lcv}
\end{figure}

The \src\ \sw/XRT light curve during the new outburst detected on 2008-09-21
is reported in Fig.~\ref{lsfig:src_lcv} in two energy ranges, together with their hardness ratio.
Since the hardness ratio was quite constant along the XRT/WT observation, we extracted a total 
spectrum.
It resulted in a net exposure time of 1159~s with an average count rate of 28.3$\pm{0.2}$~s$^{-1}$.
The fit to the 0.7--10 keV WT spectrum with an absorbed  power law model is unacceptable ($\chi^{2}_{\nu}$=1.309 for 630 dof).
A significantly better fit can be obtained either with a cutoff power law ($\chi^{2}_{\nu}$=1.187 for 629 dof)
or with a power law model together with a blackbody ($\chi^{2}_{\nu}$=1.116 for 628 dof).
We note that an absorbed blackbody model is the worst fit to the WT data, resulting in a reduced $\chi^{2}_{\nu}$$>$3.6.
The peak flux of $\sim$2.5$\times$$10^{-9}$~erg~cm$^{-2}$~s$^{-1}$ translates into an X--ray luminosity
of 2.5$\times$$10^{36}$~erg~s$^{-1}$ (at 3 kpc).

The spectral parameters resulting from these fits are reported in Table~\ref{tab:src:specxrt}.
A second total spectrum from the fainter emission observed in PC mode has been also investigated,
yielding a spectrum with a net exposure of 4100~s, and a fainter rate of 0.26$\pm{0.08}$~s$^{-1}$.  
A fit with a single absorbed power law results into a softer spectrum 
than the brighter emission observed in WT mode 
(see Table~\ref{tab:src:specxrt} for the PC spectral results).

 \begin{table*}
 \begin{center}
 \caption{\src\ spectral fits of XRT data.}
 \label{tab:src:specxrt}
 \begin{tabular}{llllrrrrr}
 \hline
 \hline
 \noalign{\smallskip}
 Spectrum  &  Time & Model & $N_{\rm H}$ 	   &  $\Gamma$                 &   kT$_{\rm bb}$/E$_{\rm cut}$  &  Flux$^{\mathrm{a}}$  & R$_{\rm bb}$             & $\chi^{2}_{\nu}$/d.o.f. \\
           &  (s since trigger) &        & ($10^{22}$ cm$^{-2}$) &                           &    (keV)     &                     &   (km)$^{\mathrm{b}}$ &         \\
  \noalign{\smallskip}
 \hline
 \noalign{\smallskip}
  Total WT  & 523--1653      & Pow  &  $0.25_{-0.02}^{+0.02}$    &  $0.82_{-0.03}^{+0.03}$ &                         & 2.7  &                     & $1.309/630$\\
            &       & Pow+BB   &  $0.43_{-0.09}^{+0.09}$ &  $2.20_{-0.04}^{+0.04}$ & $1.95_{-0.08}^{+0.08}$  &  2.5  &  $1.22_{-0.12}^{+0.12}$  & $1.116/628$\\
           &       & \sc{cutoffpl}&  $0.11_{-0.03}^{+0.03}$    &    $0.22_{-0.12}^{+0.12}$ &  $6.6_{-1.1}^{+1.5}$ & 2.5  &                       & $1.187/629$\\
Total PC    &  4880--23388      & Pow  &  $0.70_{-0.19}^{+0.24}$    &  $1.10_{-0.16}^{+0.18}$ &                       & 0.056     &                     & $1.126/50$\\
            &       & BB   &  $<0.1$                  &                       & $1.44_{-0.09}^{+0.09}$  & 0.044    &  $0.31_{-0.03}^{+0.04}$ & $1.170/50$\\
\noalign{\smallskip}
  \hline
  \end{tabular}
  \end{center}
  \begin{list}{}{}
  \item[$^{\mathrm{a}}$]{Unabsorbed 1--10\,keV flux in units of $10^{-9}$ erg cm$^{-2}$ s$^{-1}$.}
  \item[$^{\mathrm{b}}$]{Assuming a distance of 3~kpc.}
  \end{list}
  \end{table*}

\begin{figure}
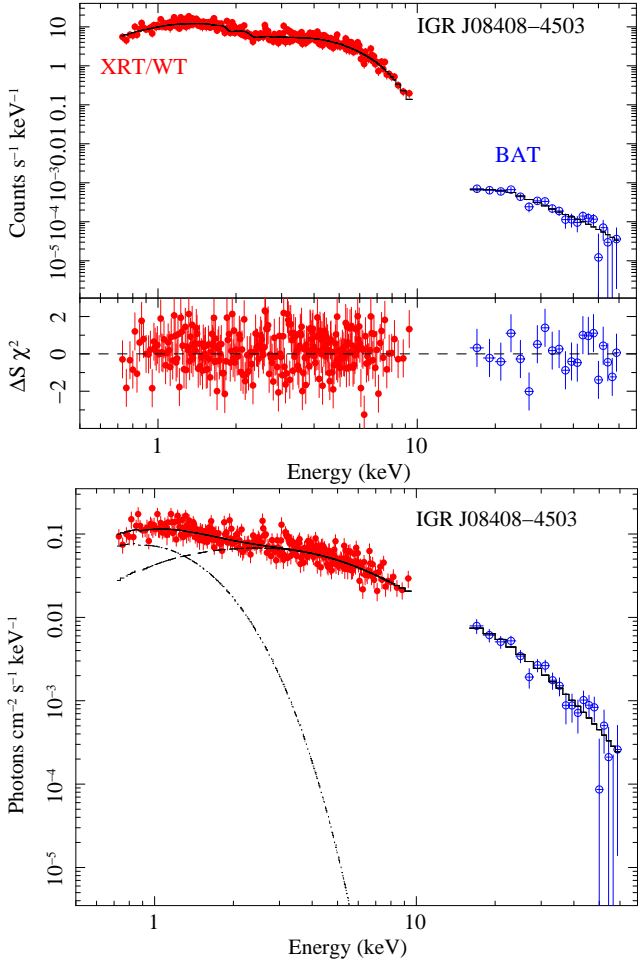

\begin{center}
\includegraphics*[angle=270,scale=0.365]{igr08040_ldadelchi_bb_bmc_highecut_new.ps}
\vspace{0.5cm}
\includegraphics*[angle=270,scale=0.35]{igr08408_ufs_bb_bmc_highecut_new.ps}
\end{center}
\caption{\sw/XRT broad band spectrum (BAT plus simultaneous WT data) of \src, fitted with a 
{\sc bmc} model in {\sc xspec} modified with a high energy cutoff ({\sc highecut} in {\sc xspec}),
together with a simple blackbody at low energies (last model in Table~{tab:src:spec}).
The {\it upper panel} shows the counts spectrum together
with the residuals in units of standard deviations. The {\it lower panel} shows the photon spectrum.
}
\label{lsfig:src:spec}
\end{figure}

A BAT spectrum could also be extracted at the beginning of the observation, with a net exposure time
of 200~s and a source count rate of 35.8$\pm{0.5}$~s$^{-1}$ (14--60 keV). 
Fitting the BAT and XRT simultaneus spectra with a single power law resulted in
an unacceptable reduced $\chi^{2}_{\nu}$ of 2.38 (293 dof). 
A very good fit can be obtained with a cutoff power law 
({\sc cutoffpl} in {\sc xspec}; $\chi^{2}_{\nu}$=1.029, for 292 dof),
obtaining  a hard spectrum with a 
photon index of 0.5, a cutoff at 13~keV, and a luminosity of 10$^{37}$~erg~s$^{-1}$ at 3~kpc.
We also tried other deconvolutions of the broad band spectrum, 
i.e. Comptonization emission models (thermal and with bulk motion) or a power law plus a blackbody model,
but they always yielded unacceptable fits with structured residuals at high energies.
We next adopted more complex models for the continuum, as a {\sc bmc} model modified with a high energy cutoff.
This resulted in a better fit than the single absorbed {\sc bmc} model, but structured residuals still appear
below 10~keV. 
The best fit is obtained adding a blackbody model to a  {\sc bmc} modified with a 
high energy cutoff ({\sc highecut} in {\sc xspec}).
A summary of all the spectral parameters reported in Table~\ref{tab:src:spec}, 
while the best deconvolution of
the broad band \src\ emission 
is shown in Fig.~\ref{lsfig:src:spec}.

Unlike \xte,  a cut-off is clearly needed in the spectrum of \src. 
Indeed {\sc bmc} alone (that is a non-attenuated power-law at high 
energies) does not fit the data well and the multiplicative factor 
{\sc highecut} is needed. We note that the fit clearly points to two distinct photon 
populations ($\sim$0.3\,keV and 1.5-2\,keV) but the current statistics 
does not allow us to constrain which of the two is seen directly as a 
blackbody and which provides part of the seed photons for Comptonization 
(hence we include twice the same model in Table~\ref{lsfig:src:spec}, one per 
configuration. See the Discussion section for possibile interpretations).

 \begin{table*}
 \begin{center}
 \caption{Spectral fits of simultaneous XRT and BAT data of \src. See Section~\ref{bmc} for the description of the adopted models.}
 \label{tab:src:spec}
 \begin{tabular}{lrrrrrrrrrr}
 \hline
 \hline
 \noalign{\smallskip}
Model  &  	   &         &       &   Parameters &          &                        &      &  &     \\
 \hline
\sc{cutoffpl}     &  $N_{\rm H}$$^{\mathrm{a}}$   &     $\Gamma$          &   E$_{\rm cut}$$^{\mathrm{b}}$  &  & & & &  &  Flux$^{\mathrm{c}}$  & $\chi^{2}_{\nu}$/d.o.f. \\
                  &  $0.10_{-0.04}^{+0.04}$      &  $0.50_{-0.08}^{+0.08}$ &     $13_{-2}^{+2}$         &  & &  & & &    10             & $1.029/292$\\

\sc{bmc*highecut+bb} &  $N_{\rm H}$$^{\mathrm{a}}$ &    kT$_{\rm BB}$$^{\mathrm{b, d}}$     &   $\alpha$$^{\mathrm{d}}$   & $\log(A)$$^{\mathrm{d}}$ &  E$_{\rm cut}$$^{\mathrm{b}}$    & E$_{\rm fold}$$^{\mathrm{b}}$  & kT$_{\rm BBody}$$^{\mathrm{b}}$  & R$_{\rm BBody}$$^{\mathrm{e}}$ & Flux$^{\mathrm{c}}$& $\chi^{2}_{\nu}$/d.o.f. \\

 &  $0.10_{-0.05}^{+0.70}$ &  $0.32_{-0.06}^{+0.05}$ & $6_{-5}^{+13}$$\times10^{-4}$ & $3.3_{-0.2}^{+0.5}$ & $21_{-7}^{+4}$ & $18_{-4}^{+5}$ & $1.85_{-0.13}^{+0.16}$ & $1.2_{-0.8}^{+0.2}$  & 10 & $0.996/287$\\

\sc{bmc*highecut+bb} &  $N_{\rm H}$$^{\mathrm{a}}$ &    kT$_{\rm BB}$$^{\mathrm{b, d}}$     &   $\alpha$$^{\mathrm{d}}$   & $\log(A)$$^{\mathrm{d}}$ &  E$_{\rm cut}$$^{\mathrm{b}}$    & E$_{\rm fold}$$^{\mathrm{b}}$  & kT$_{\rm BBody}$$^{\mathrm{b}}$  & R$_{\rm BBody}$$^{\mathrm{e}}$ & Flux$^{\mathrm{c}}$& $\chi^{2}_{\nu}$/d.o.f. \\

 &  $0.10_{-0.03}^{+0.06}$ &  $1.4_{-0.1}^{+0.1}$ & $0.35_{-0.06}^{+0.08}$ & $0.82_{-0.14}^{+0.15}$ & $21_{-12}^{+5}$ & $22_{-5}^{+8}$ & $0.33_{-0.04}^{+0.82}$ & $12_{-1}^{+4}$  & 10 & $0.988/287$\\
\noalign{\smallskip}
  \hline
  \end{tabular}
  \end{center}
  \begin{list}{}{} 
 \item[$^{\mathrm{a}}$]{Absorbing column density is in units of $10^{22}$ cm$^{-2}$.}
   \item[$^{\mathrm{b}}$]{High energy cutoff (E$_{\rm cut}$), e-folding energy (E$_{\rm fold}$) and blackbody temperatures are in units of keV.}
  \item[$^{\mathrm{c}}$]{Unabsorbed 0.1--100\,keV flux is in units of $10^{-9}$ erg cm$^{-2}$ s$^{-1}$.}
 \item[$^{\mathrm{d}}$]{kT$_{\rm BB}$ is the blackbody color temperature of the seed photons in the {\sc bmc} model, $\alpha$ 
is the spectral index and Log(A) is the illumination parameter.}
\item[$^{\mathrm{e}}$]{The radius of the {\sc bbody} model is in units of km at an assumed source distance of 3 kpc.}
  \end{list}
  \end{table*}

\subsection{Timing analysis}

We performed a timing analysis on the three sources to
investigate for the presence of X-ray pulsations.
A $Z^2 _1$ test \citep{Buccheri1983} on the fundamental harmonics was
applied  on the events collected in each WT mode sequence (see Table~1) 
searching in  the frequency range between 0.005 and 100 Hz with a
frequency resolution of 1/$\Delta$T Hz, where $\Delta$T is the duration of the WT segment.
The resulting power spectrum does not reveal any significant
deviations from a statistically flat distribution. Data collected in PC mode
were not analyzed because of their much lower statistics content.

\section{Discussion and Conclusions}

We report on three new outbursts from three different Supergiant Fast X--ray Transients,
\xte, \igr, and \src, observed with \sw.

\subsection{Spectroscopy}

All these three sources were observed in outburst  with \sw\ in the past, thus allowing
a proper comparison between the spectral properties of the different flares.

For \igr\  only the spectrum below 10 keV is available. 
Compared with the emission previously observed (Paper~III)
it displays a more absorbed and softer emission (single power law model).
A similar behaviour from \igr\ has been recently reported by \citet{Rampy2009} during
a Suzaku observation catching the source during a long outburst (at least three days of accreting phase) in March 2008
(the same reported in Paper~III). The XRT/WT spectrum is compatible with spectral parameters
reported for the segment n.6 of the XIS observation \citep{Rampy2009}. 
A good fit to the \igr\ XRT spectrum can  also be obtained with a black body model,
resulting in a temperature of 1--2~keV, and in a black body radius (at 3.6~kpc) of about 1--1.5~km,
compatible with an origin in the neutron star polar cap.

\xte\ displays for the first time a variable absorption column density within a flare.
This behaviour  was also observed in \src\ during the multiple flaring
activity caught by \sw\ in July 2008 \citep{Romano2009:sfxts_paper08408}.
A higher absorption is observed during the rise to the flare peak, so it is possibly
due to the accumulation of matter onto the compact object, instead of
obscuring material into the line of sight located far away from the central X--ray source.
The average column density is intermediate between the out-of-outburst emission (Paper~I)
and that displayed during the previous bright flare reported in Paper~III. 
No correlation of the spectral parameters with the source flux can be found.
A variability on short timescales 
in the obscuring column density of a similar amount was observed also during an outburst from \igr\ 
with Suzaku (Rampy et al. 2009). This sudden absorption episode was interpreted as 
due to the transit of a foreground dense cloud of matter passing in front of the compact object, and as 
the first direct evidence at X-rays for dense clumps of matter in the supergiant wind.
The simultaneous spectroscopy of the XRT and BAT data of \xte\  can be 
adequately performed using a simple {\sc bmc} model (Table~\ref{tab:xte:spec}): a significant fraction 
($\log A>0$) of the initial black body (BB) seed photons ($kT_{\rm BB}\sim1.6$\,keV) is 
efficiently  up-scattered ($\alpha\sim$1.3) by the Comptionizing plasma. 
The radius associated with the BB component of the {\sc bmc} model 
is about 1.6~km (at 2.7 kpc), thus consistent with
a polar cap origin.
The model {\sc compTT} allow us to quantify the physical conditions of the Comptonizing plasma,
since it returns the plasma temperature $kT_{e}$ and 
optical depth $\tau$, instead of the Comptonization efficiency $\alpha$. 
As shown in Table~3, the $kT_{e}$ temperature  could not be 
constrained and this is consistent with the fact that the data could be fit well 
by the {\sc bmc} model that indeed has no cut-off in its spectral shape. These 
two results point to the fact that the current statistics and data 
coverage  \emph{do not require} a cutoff, although they cannot 
exclude it. 
The reason why the {\sc cutoffpl} model fits the data much better than a simple 
power law resides in the fact that the residuals using a simple power law show an 
excess around 1--2\,keV, whereas the curved shape of the 
{\sc cutoffpl} linked to the interplay with the absorbing column density 
can describe the data in a satisfactory way. 
We note that this excess is naturally taken into account 
in the physical models, {\sc compTT} and {\sc bmc}, where a 1--2\,keV BB seed photon 
population is obtained. 
Unfortunately the current data-set does not allow us to investigate the 
evolution of the high energy part of the spectrum, so little can be 
currently said  about the possible evolution of the Comptonizing medium. 
For the remaining  part of the outburst, only the softer part below 
10\,keV is available. Nevertheless, it can be seen from Table~2 that a BB  
is more suitable to fit the \sw/XRT data rather than a power law, 
consistent with what is obtained in the overall XRT+BAT spectrum. 

A comparison of the \src\ broad band spectrum extracted from the bright flare observed in September 2008
with the ones observed with \sw\ in July 2008 and in October 2006 \citep{Romano2009:sfxts_paper08408}
reveals that the new spectrum is more similar to that observed in 2006 (very low absorption,
a hard photon index, and a similar high energy cutoff at around 10--15 keV).
Unlike \xte, the spectrum of \src\ could not be fit 
with a single Comptonization model, two additive components were needed: 
a blackbody plus a {\sc bmc}  model (the latter with high energy cutoff).
This could be due to the very low absorption at low energy
with respect to the other SFXTs studied here that required 
an additional component to take into account for the softer part of the spectrum. 
The presence of the cutoff implies that the overall spectrum
is the result of thermal Comptonization (the {\sc bmc} model alone 
has a non-attenuated power law shape). The low value of the 
$\alpha$ parameter obtained ($\alpha$$<$0.4, $\Gamma<1.4$) is typical 
for saturated Comptonization.

The seed photon temperature for the thermal Comptonization {\sc bmc} component 
and the BB temperature could not be linked to the same value in 
the fitting process and indeed resulted in two clearly 
different photon populations, a cold one at about 0.3\,keV 
and a hotter one at 1.4--1.8\,keV. The current data 
did not allow us to establish in a solid way, which one 
of these two populations is seen directly as a BB and which one 
ends up being seed photons for the thermal Comptonization. 
As can be seen in Table~\ref{tab:src:spec} both scenarii are statistically 
acceptable. In one case we obtain a cold (0.3\,keV) BB 
of about R=12\,km directly visible (few percent of the total flux), 
together with a hotter photon population (1.4\,keV) thermally Comptonized 
(the dominant component), part of which is directly visible (logA=0.8). 
This could depict a thermal halo around the neutron star [0.3\,keV, as in \citet{Ferrigno2009}] 
together with BB seed photons from the accreting material, 
part of which is directly visible (e.g at the column boundaries) and part 
is thermally Comptonized (from the accreting matter). 
This scenario is consistent with what observed for \xte\ (Table~3) 
with the thermal cold halo buried in the high column absorption (an 
order of magnitude higher than for \src).    

In the second case we obtain a hotter BB  (1.9\,keV, possibly from 
the base of the accreting column, R$\sim$1\,km) directly visible, 
accounting for about 30\% of the total flux, together with a colder 
plasma (0.3\,keV) embedded in a thermally Comptonizing medium 
(logA$\sim$3) such as an atmosphere confined by multi-polar or crustal 
components of the magnetic field [e.g. \citet{Ferrigno2009} and references 
therein]. 

With the information at hand, we cannot exclude either of these scenarii.
The spectra of HMXBs have been generally described by phenomenological 
models and this work is one of the few cases where two distinct spectral 
components linked to two different physical conditions have been observed 
[see also \citet{Ferrigno2009}].

\subsection{Search for periodicities}

\igr\ was previously observed in outburst with \sw\ in two occasions:
on 2007 November 8 and on 2008 March 31 (Paper~III). 
This implies that the three bright flares are spaced by $\sim$144 and $\sim$157~days, respectively.
We note however that in \igr\ the flare occurred on 2008 September 4
was preceded by intense activity in the previous few days during the observations part of 
the INTEGRAL Galactic bulge monitoring program 
performed on 2008, August 30, reaching a flux of about 50 mCrab (18--40 keV, \citealt{Romano2008:Atel1697}).
This seems to indicate an outburst phase which began a few days before the BAT trigger,
suggesting an outburst duration of several days  (an outburst lasting at least 3~days has also been caught 
with Suzaku \citep{Rampy2009}): this could imply a periodic occurrence of the
bright flaring activity, every $\sim$150~days.
If the X--ray bright flares are
triggered periodically during the periastron passage in an eccentric binary,  the
orbital period is probably about 150~days in \igr.
This is consistent with previous findings with INTEGRAL, where a possible outburst recurrence timescale of 
165$\pm{3}$~days has been suggested \citep{Walter2006}.

From the times of previous \src\ flares, we suggested  \citep{Romano2009:sfxts_paper08408} that
a double-periodicity outburst recurrence of $\sim$11~days and 24~days was present, thus consistent
with the picture where the outbursts are triggered when the compact object, along its orbit, 
crosses twice an inclined second component of an outflowing dense wind, confined along a preferential plane 
(e.g. the supergiant equator), inclined with respect to the orbital plane. 
On the other hand, the last flare from \src\ 
did not occur at the right times predicted by these double periodicities 
(the nearest outburst was predicted to occur on 2008, September 13,
instead of 2008 September 21).
This could indicate that either these derived periodicities are actually wrong (and the flaring
activity is not periodic but sporadic), or that another
mechanism is at work when producing this latter kind of outburst: 
a possible explanation is that, while the previous outbursts were produced when
the neutron star crossed twice the denser wind component, 
this latter outburst was triggered 
when the neutron star approached the periastron passage, 
accreting matter from the polar wind, in an eccentric orbit. 

The three sources analysed here do not show any evidence for X--ray pulsations (see Sect.~3.4).

\subsection{SFXTs as a class}

\begin{figure}
\begin{center}
\centerline{\includegraphics[width=9cm,angle=0]{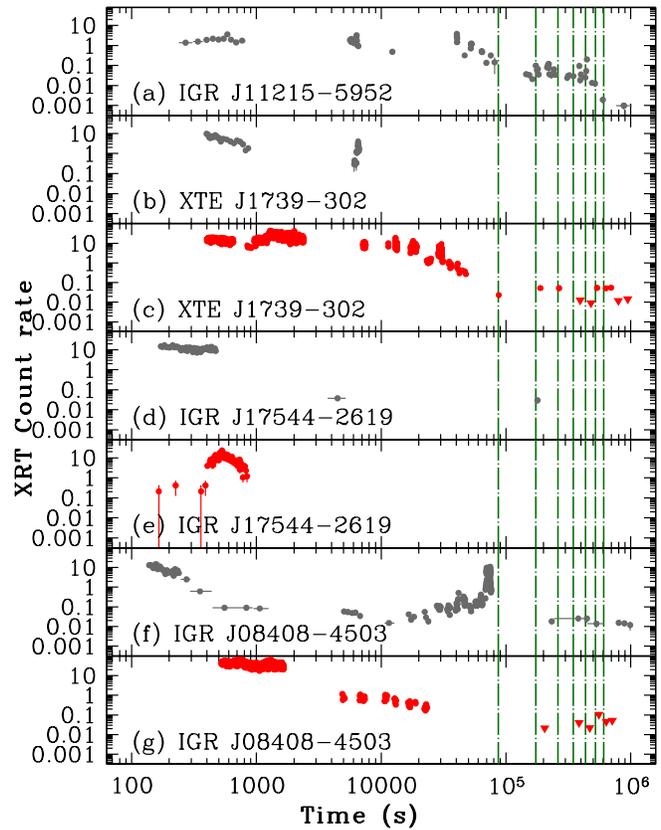}}
\caption{Light curves of the outbursts of SFXTs followed by {\it Swift}/XRT
referred to their respective BAT triggers. Points denote detections, triangles 3$\sigma$ upper limits.
Red data points (panels c, e, g) refer to observations presented here for the first time, 
while grey data points refer to data presented elsewhere.  
Note that where no data are plotted, no data were collected.  
Vertical dashed lines mark time intervals equal to 1 day, up to a week. 
We show the IGR~J11215$-$5952 light curve [panel (a)] with an arbitrary start time, since
the source did not trigger the BAT \citep{Romano2007}.
The panels (b) and (c) report the two flares from  XTE~J1739--302 observed
on 2008 April 8 (Paper~III) and on 2008 August 13 (reported here), respectively.
The panels (d) and (e) show the outbursts of IGR~J17544--2916 which occurred on 2008 March 31 
(Paper~III) and the latest one reported here (this outburst, as in IGR~J11215$-$5952, 
did not trigger the BAT, thus the start time is arbitrary and was set
at the beginning of the observation).
The panels (f) and (g) show the last two outbursts from 
another SFXT, not part of the XRT campaign, IGR~J08408--4503, which 
occurred on 2008 July 5 \citep{Romano2009:sfxts_paper08408} 
and on 2008, September 21 (reported in this paper).
	}
\label{lsfig:duration}
\end{center}
\end{figure}

During the last outburst, a very good sampling of the \xte\ light curve was possible (the best light curve 
during an outburst from this source, to date; see Fig.~\ref{lsfig:duration}c).  
We can compare the \emph{Swift}/XRT light curve of XTE~J1739--302
with the X-ray luminosity predicted by a Bondi-Hoyle accretion
onto the neutron star from a spherically symmetric homogeneus wind
for different values of the orbital period and eccentricity.
We assume for the supergiant a stellar mass of 33~M$_{\odot}$,
a radius of 23~R$_{\odot}$ \citep{Vacca1996},
a beta-law for the supergiant wind with $\beta=1$,
a wind terminal velocity of 1900~km~s$^{-1}$,
a wind mass loss rate of 10$^{-6}$~M$_{\odot}$~yr$^{-1}$
and a temperature of the stellar wind of 10$^5$~K.
We find that for any choice of the orbital period and eccentricity
the decline of the light curve observed with \sw\
is too rapid compared with the calculated light curve,
even adopting high values for the orbital period and eccentricity.
Since a spherical distribution of wind matter is not able to
explain the observed shape of the X--ray light curve, a possible explanation is
the presence of a second outflowing wind component,
denser than the supergiant polar wind, which
is crossed by the neutron star along its orbit, in analogy to what we proposed  
for IGR~J11215--5952 \citep{Sidoli2007}.
Alternatively, if we consider the clumpy wind scenario, 
from the duration of the bright part of the flare in \xte\ ($>$0.6~days) and 
its luminosity ($\sim$10$^{36}$~erg~s$^{-1}$), we can derive an accreted mass 
of $>$4$\times$10$^{25}$~g and a size of $>$10$^{13}$~cm \citep{Walter2007},
which corresponds to more than 6 supergiant radii, thus making 
very unlikely that it is a very large single clump ejected by the supergiant \citep{Walter2007}.
Instead, it could be 
alternatively explained with a huge gas stream 
composed by several smaller clumps.

In Fig.~\ref{lsfig:duration} we compare some of the outbursts from SFXTs as observed
during our monitoring campaign. In particular the three outbursts discussed here are shown in panels 
c, e and g from \xte, \igr\ and \src, respectively.
One could be tempted to conclude, from this comparison, that
different types of outburst are present, even in the same source. 
It is actually not possible to compare, for example, the two outbursts from \src\ (last panels in Fig.~\ref{lsfig:duration}):
the several upper limits to the flux in the declining part of the outbursts are
compatible with the source detections during the previous \src\ outburst (Fig.~\ref{lsfig:duration}f), thus
it is not possible to conclude that the new outburst from \src\ (Fig.~\ref{lsfig:duration}g) 
is shorter than the previous one (Fig.~\ref{lsfig:duration}f).
On the other hand, the  light curve from \xte\ (Fig.~\ref{lsfig:duration}c) allows us 
to perform a proper comparison with the
outburst from the periodic SFXTs IGR~J11215--5952 (Fig.~\ref{lsfig:duration}a; 
\citealt{SidoliPM2006}, \citealt{Romano2007}):
these two sources appear to undergo similar outbursts with a similar duration.
Interestingly, Rampy et al. (2009) report on a long outburst activity from \igr\ during the March 2008 outburst,
with a rise time much longer than what observed with Chandra during an X-ray flare in 2004 \citep{zand2005} in the
same source. These authors suggest that different kinds of outbursts can occur in the same SFXT.

From the duration of single well sampled short flares in \src\ we derived an orbital period of about 35~days
\citep{Romano2009:sfxts_paper08408} adopting an expansion law for the clump sizes as the clump is accelerated
far away from the supergiant donor, in the framework of an inhomogeneous wind \citep{Romano2009:sfxts_paper08408}.
Adopting this same expansion law, and from the observed durations of the short flares which compose the outburst light curves 
in \xte\ and \igr, we can derive the 
distance of the accreted clump from the supergiant star, as we did in \src\ \citep{Romano2009:sfxts_paper08408}.
More details will be reported in Ducci et al. (2009, in preparation).
Fitting with a Gaussian a few well sampled flares in \xte\ we obtain a FWHM of 260$\pm{60}$~s and 390$\pm{60}$~s,
while in \igr\ the observed flare has a duration of 220$\pm{10}$~s (FWHM). 
From the clump expansion law 
reported in \citet{Romano2009:sfxts_paper08408} we derive a distance of the compact object 
from the supergiant donor in these two sources as follows:
in the range from 6.8$\times$10$^{12}$~cm to 1.25$\times$10$^{13}$~cm for \xte, 
while near to 6.8$\times$10$^{12}$~cm for \igr\ 
(assuming  a supergiant mass of 33~M$_{\odot}$ and a stellar radius of 23~R$_{\odot}$).
Assuming a circular orbit these distances translate into orbital periods ranging 
from 20 to 50~days in \xte\ and around
20~days for \igr. 
This latter orbital period  
is not consistent with that of $\sim$150~days, suggested by the
outburst recurrence in \igr. This discrepancy can be easily reconciled if the orbit in this SFXT 
is highly eccentric.

\citet{Jain2009} recently reported on the discovery of an orbital period of 3.3~days from the SFXT 
\med. This orbital periodicity is even shorter 
than that displayed by several persistent HMXBs with supergiant companions. 
This poses serious problems to the different 
physical mechanisms proposed for SFXTs, implying an orbital separation of 
about 2$\times$10$^{12}$~cm (assuming a supergiant  mass  
of $\sim$30~M$_{\odot}$), which is about 1.2 stellar radii, 
thus well inside the region where the highest wind clumps 
number density  is expected, and a persistent X--ray emission is predicted \citep{Negueruela2008}. 
The very different orbital periods discovered in SFXTs to date (ranging from 3.3 days to 165~days) possibly
point to different kinds of SFXTs (different mechanisms at work in different sources, and possibly in the 
same source, as discussed above).

\section*{Acknowledgments}

LS thanks INAF-IASF Palermo and PR thanks INAF-IASF Milano, 
where some of the work was carried out, for their kind hospitality. 
We thank Valentina La Parola and Cristiano Guidorzi for helpful discussions.
We thank the \sw\ team for making these observations possible,
the duty scientists, and science planners. 
This work was supported in Italy by ASI contracts  I/023/05/0, 
I/088/06/0 and I/008/07/0,
and partially by the grant from PRIN-INAF 2007, ``Bulk motion Comptonization models in X-ray 
Binaries: from phenomenology to physics'' (PI M. Cocchi).
This work  was supported at PSU by NASA contract NAS5-00136. 
HAK was supported by the {\it Swift } project. 
DNB and JAK acknowledge support from NASA contract NAS5-00136.


\bsp
\label{lastpage}
\end{document}